\documentstyle[preprint,aps]{revtex}

\tightenlines
\input{epsfig.sty}
\begin{document}

\preprint{}


\title{Spatial structure of quark Cooper pairs 
in a color superconductor}

\author{Masayuki Matsuzaki
 \thanks{E-mail address: matsuza@fukuoka-edu.ac.jp}
}

\address{Department of Physics, Fukuoka University of Education, 
Munakata, Fukuoka 811-4192, Japan}

\date{\today}
\maketitle

\begin{abstract}
Spatial structure of Cooper pairs with quantum numbers color {\bf3}$^*$, 
$I=J=L=S=0$ in $ud$ 2 flavor quark matter is studied by solving the 
gap equation and calculating the coherence length in full momentum range 
without the weak coupling approximation. Although the gap at the Fermi 
surface and the coherence length depend on density weakly, the shape 
of the $r$-space pair wave function varies strongly with density. This result 
indicates that quark Cooper pairs become more bosonic at higher 
densities.
\end{abstract}


  Color superconducting phases of strongly interacting matter at 
high density and low temperature \cite{fd1,fd2,fd3} are attracting much 
attention recently. They were studied first as an example of 
pair condensation in relativistic many-body systems by Bailin 
and Love in the early 80's \cite{bl} (see also Ref.\cite{bar}). 
They also mentioned the 
nucleon-nucleon pairing as another example; its detailed study 
was begun in the early 90's \cite{kr} and is developing 
recently \cite{rel1,rel2,mm,tm}. A study of the quark-quark pairing in 
an SU(2) color model was done at almost the same 
time \cite{su21} (see also Refs.\cite{su22,lat}). Iwasaki and Iwado's 
work on a 1 flavor system \cite{ii} was the first study of the 
realistic SU(3) color system  (see also Ref.\cite{i}). Since the works 
of Rapp {\it et al.} \cite{inst1} and Alford {\it et al.} \cite{inst2} 
on $ud$ 2 flavor system, color superconductivity has been 
studied extensively. The $^1S_0$ ($J=L=S=0$) state in the 2 
flavor case \cite{hh} and the color-flavor-locked state in 
$uds$ 3 flavor case \cite{csc19,csc20,csc21,csc22,csc23}, 
respectively, have been understood 
to be the most favored channels. The magnetic 
interaction has been shown to be responsible  for these pair 
condensations \cite{csc24,csc25,csc26,csc27,csc28,csc29}. 
Their astrophysical consequences were 
also studied \cite{blas1,blas2}. 

  The purpose of this Brief Report is to visualize the spatial 
structure, in particular its density dependence, of quark 
Cooper pairs, which has not been discussed, to the author's 
knowledge. This is done for Cooper pairs of the simplest form; 
color ${\bf 3}^{\ast}$, isosinglet $^1S_0$ pairs in the 
2 flavor system. To this end, calculations are performed 
for zero temperature under an instantaneous approximation 
while full ${\bf k}$-dependence is retained. That is, 
we allow strong coupling in the sense that momenta far from 
the Fermi surface also contribute.
We will mention briefly the frequency dependence later.

  We start from deriving the gap equation for $\Delta(k)$ 
representing the gap for color ${\bf 3}^{\ast}$, 
isosinglet $^1S_0$ pairs composed of two quarks which are 
time-reversal conjugate to each other with respect to space and spin, 
\begin{eqnarray}
&&\Delta_{{\bf k}sfi,{\bf k}'s'f'i'}
  =(-1)^{\frac{1}{2}-s}\delta_{{\bf k},-{\bf k}'}
   \delta_{s,-s'}\epsilon_{ff'}\hat\epsilon_{ii'}
   \Delta(k)\, , \nonumber \\
&&\,\,\epsilon_{ff'}=\left(
\begin{array}{@{\,}cc@{\,}}
              0 & 1 \\
             -1 & 0
\end{array}
\right)\, , \,\,
\hat\epsilon_{ii'}=\left(
\begin{array}{@{\,}ccc@{\,}}
              0 &  1 & -1 \\
             -1 &  0 &  1 \\
              1 & -1 &  0
\end{array}
\right)\, ,
\end{eqnarray}
where $s$, $f$ and $i$ denote spin, flavor and color, 
respectively. Note that time-reversal is represented by 
$T$ times complex conjugate, and $T=-iC\gamma^5$. 
We work in a Hamiltonian formalism since it is more 
convenient than the Gor'kov formalism \cite{gor} in the present 
study in which only the pair condensate is considered. 
The latter might be more suitable when the coupling between 
the pair condensate and the chiral condensate \cite{inst2} 
is considered. Such a formalism has already been developed for the 
nuclear system \cite{rel1,mm} although the $N$-$\bar N$ condensate 
is negligibly small \cite{mm} due to a large nucleon mass. 
The resulting $12\times12$ Hamiltonian matrix,
\begin{eqnarray}
&&\left(
\begin{array}{@{\,}cc@{\,}}
              E_k-\mu      & \Delta \\
              -\Delta^\ast & -(E_k-\mu)
\end{array}
\right)\, , \nonumber \\
&&\,\,E_k=\sqrt{{\bf k}^2+M_q^2}, \,\, \mu=E_{k_{\rm F}}\, ,
\end{eqnarray}
is easily block-diagonalized to two $6\times6$ matrices 
by inspection. Then they are fully diagonalized along the 
lines of the 1 flavor case studied in Ref.\cite{ii}. 
By expressing the pair condensate in terms of the Bogoliubov 
coefficients, we obtain 
\begin{eqnarray}
&&\Delta(p)
=-\frac{1}{8\pi^2}  \int_0^\infty \bar v(p,k)
  \frac{\Delta(k)}{E'(k)}k^2dk\, ,
\nonumber \\
&&\,\,E'(k)=\sqrt{(E_k-E_{k_{\rm F}})^2+3\Delta^2(k)} \, .
\end{eqnarray}

  As for the residual interaction, we adopt the one gluon 
exchange interaction with the leading order screening.
Although the gluon propagator therein contains a gauge-dependent 
term in general, it vanishes due to the equation of motion of the 
external quark \cite{csc27}. Then we proceed to a static (instantaneous) 
approximation \cite{ii,i,hh}. 
At this step, the dynamic magnetic screening drops. 
By performing a spin average and an angle integration to project 
out the $S$-wave component, we obtain 
\begin{eqnarray}
&&\bar v(p,k)=-\frac{\pi}{3}\alpha_{\rm s}\frac{1}{pkE_pE_k}
\nonumber \\
&&\times\biggl(\left(2E_pE_k+2M_q^2+p^2+k^2+m_{\rm E}^2\right)
       \ln\left(\frac{(p+k)^2+m_{\rm E}^2}{(p-k)^2+m_{\rm E}^2}\right)
\biggr.
\nonumber \\
\biggl.
&&\,\,+2\left(6E_pE_k-6M_q^2-p^2-k^2\right)
       \ln \bigg|\frac{p+k}{p-k} \bigg|
\biggr)\, ,
\nonumber \\
&&\,\,\,m_{\rm E}^2=\frac{4}{\pi}\alpha_{\rm s}\mu^2\, .
\label{vpp}
\end{eqnarray}
The running coupling constant is taken from Ref.\cite{higashi} as 
in Refs.\cite{i,hh}, 
\begin{eqnarray}
&&\alpha_{\rm s}({\bf q}^2)=\frac{4\pi}{9}
   \frac{1}{\ln\left(\frac{q_{\rm max}^2+q_{\rm c}^2}
                          {\Lambda_{\rm QCD}^2}
               \right)}\, ,
\nonumber \\
&&{\bf q}={\bf p}-{\bf k}\, ,\,\, 
  q_{\rm max}=\max\{p,k\}\, ,
\end{eqnarray}
with parameters $q_{\rm c}^2~=~1.5\Lambda_{\rm QCD}^2$, 
$\Lambda_{\rm QCD}=$ 0.4 GeV.

  The frequency dependence of the gap, which is beyond the present 
formulation, has been shown recently to change the asymptotic 
behavior of the gap in the weak-coupling 
approaches \cite{csc24,csc26,csc28}. There frequency dependence 
appears both in the gluon propagator that determines the interaction 
and in the other part of the integrand in the 
Eliashberg equation \cite{csc24,csc26,csc23}. Only the effects of 
the former can be estimated approximately here by modifying 
the magnetic part of Eq.(\ref{vpp}) as
\begin{eqnarray*}
&&-\frac{\pi}{3}\alpha_{\rm s}\frac{1}{pkE_pE_k} 
\\
&&\times\left(6E_pE_k-6M_q^2-p^2-k^2-m_{\rm M}^2\right)
\ln\left(\frac{(p+k)^2+m_{\rm M}^2}{(p-k)^2+m_{\rm M}^2}\right)\, , 
\\
&&m_{\rm M}^2=\big(\vert q^0\vert\alpha_{\rm s}\mu^2\big)^{2/3}\, ,
\end{eqnarray*}
where $m_{\rm M}^2$ is originating from the Landau 
damping \cite{csc26,csc27} and the frequency $q^0$ is an independent 
parameter here due to the 
nature of the present formulation. Evidently this simply weakens 
the magnetic attraction and accordingly monotonically reduces 
the gap at the Fermi surface, 
$\Delta(q^0=10~{\rm MeV})/\Delta(q^0=0)=$ 0.70,
$\Delta(q^0=50~{\rm MeV})/\Delta(q^0=0)=$ 0.53,
at $k_{\rm F}=2$ fm$^{-1}$, for example, while the $q^0$-dependence 
of the coherence length (defined later) is rather weak.
In the Eliashberg equation the gap is given as a function of $p^0$, 
where $q^0=p^0-k^0$ and the equation includes the integration 
with respect to $k^0$. Its $p^0$-dependence was numerically 
studied in Refs.\cite{csc26,csc23}, where the gap exhibits a 
maximum at intermediate densities whereas decreases monotonically 
at asymptotically high densities.

\begin{figure}[t]
\begin{center}
\epsfig{figure=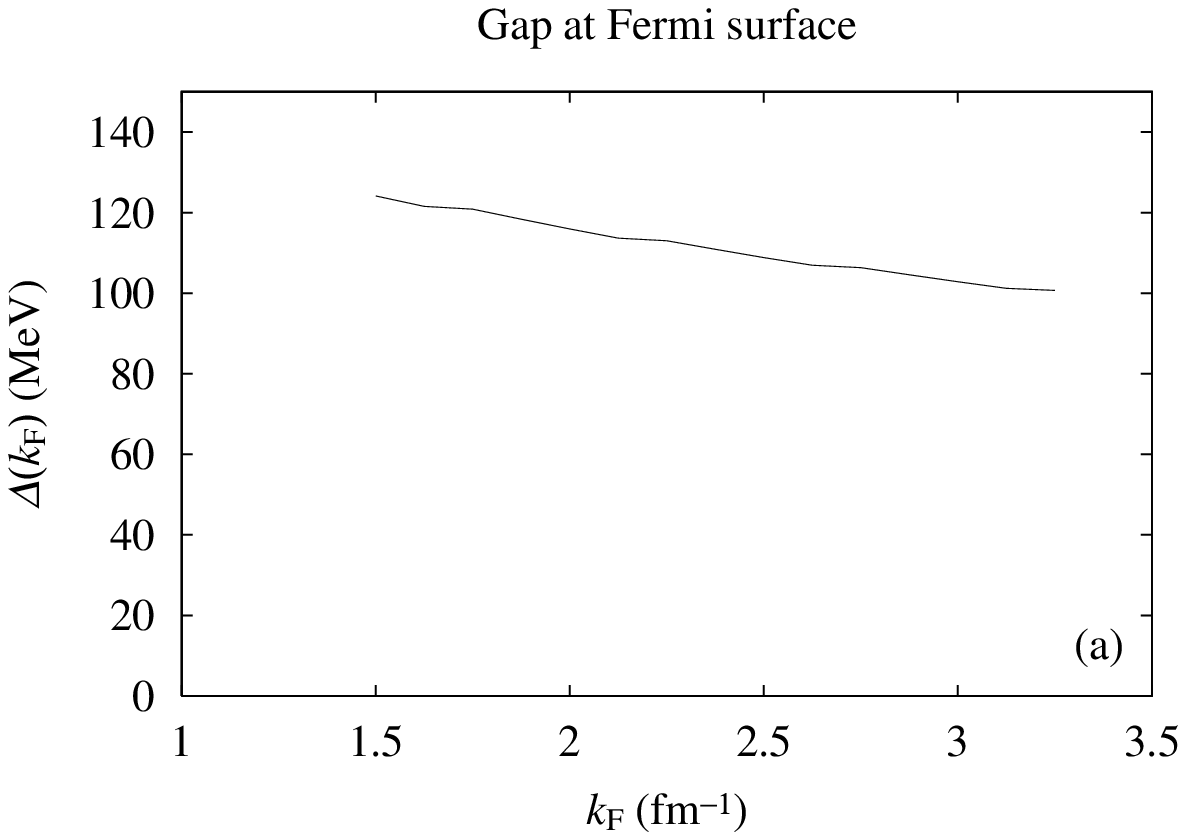,width=10cm}
\vskip 0.5cm
\epsfig{figure=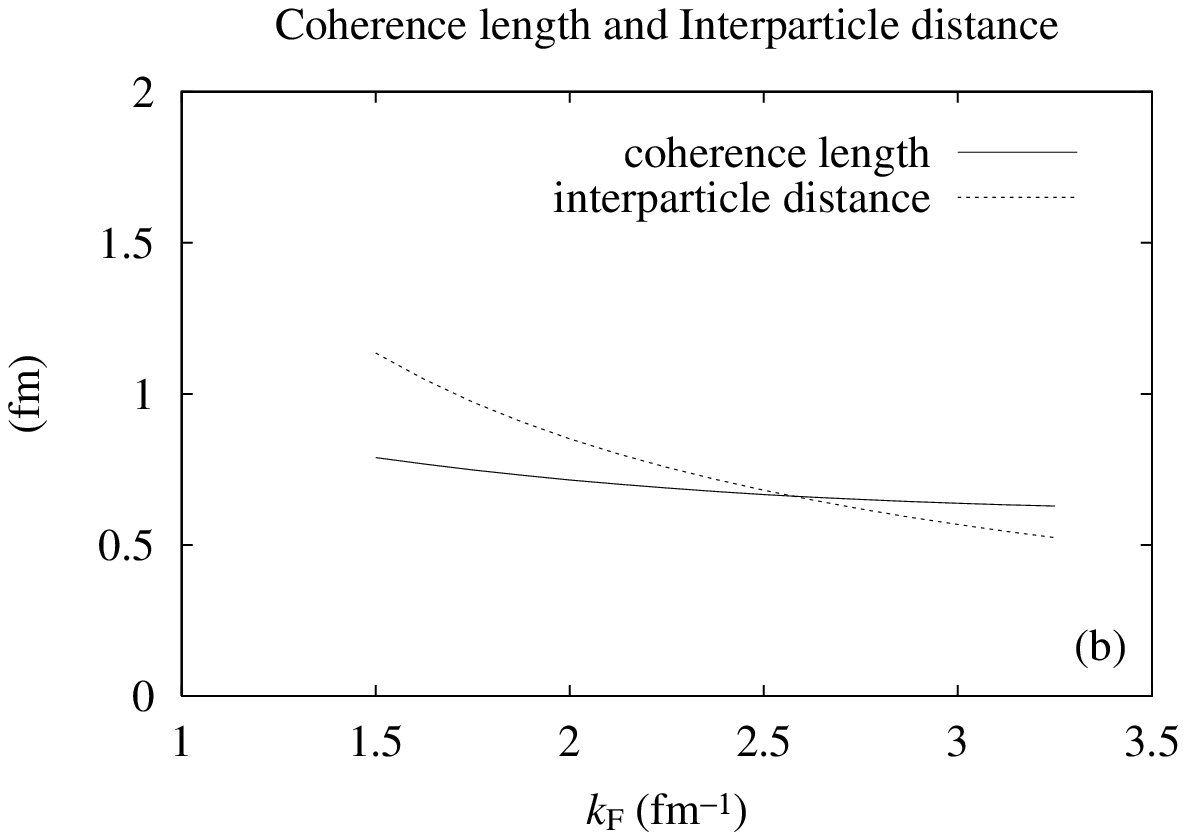,width=10cm}
\end{center}
\caption{
(a) Pairing gap at the Fermi surface, 
and (b) coherence length and average interquark distance, 
as functions of the Fermi momentum $k_{\rm F}$.
}
\label{fig1}
\end{figure}

\begin{figure}[t]
\begin{center}
\epsfig{figure=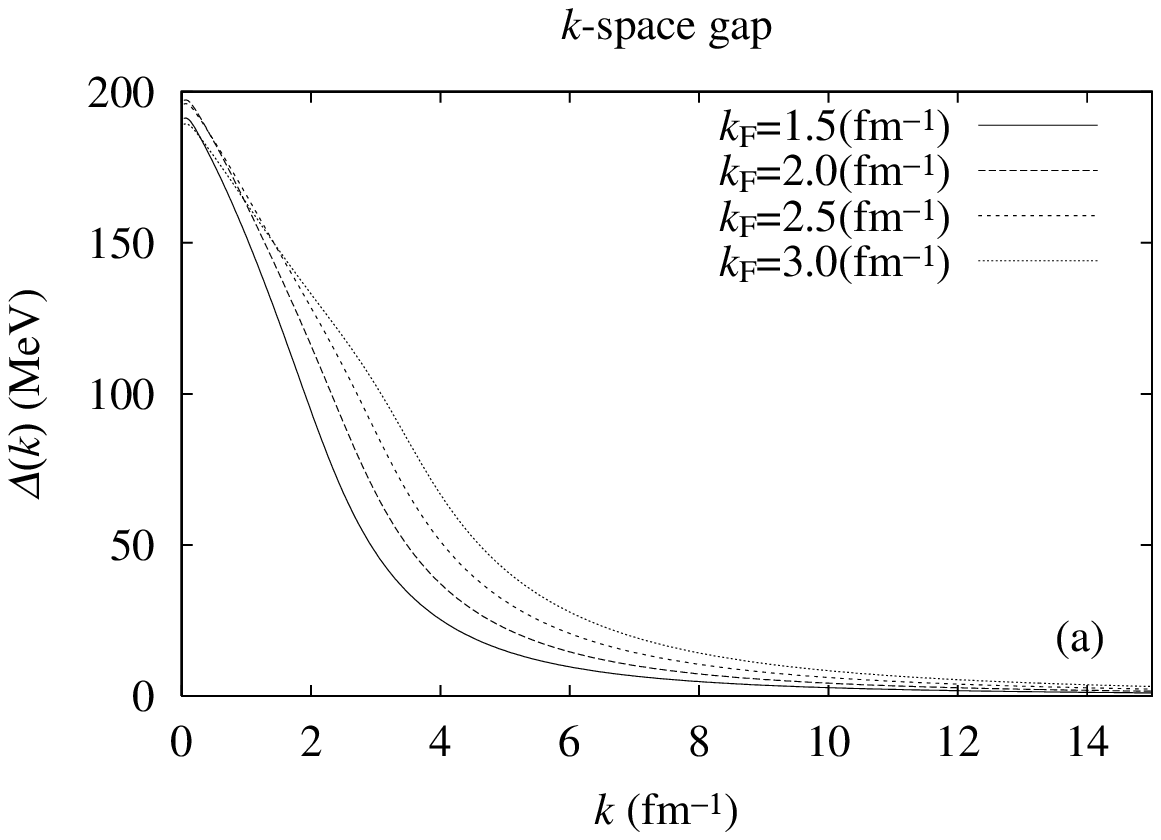,width=8cm}
\vskip 0.5cm
\epsfig{figure=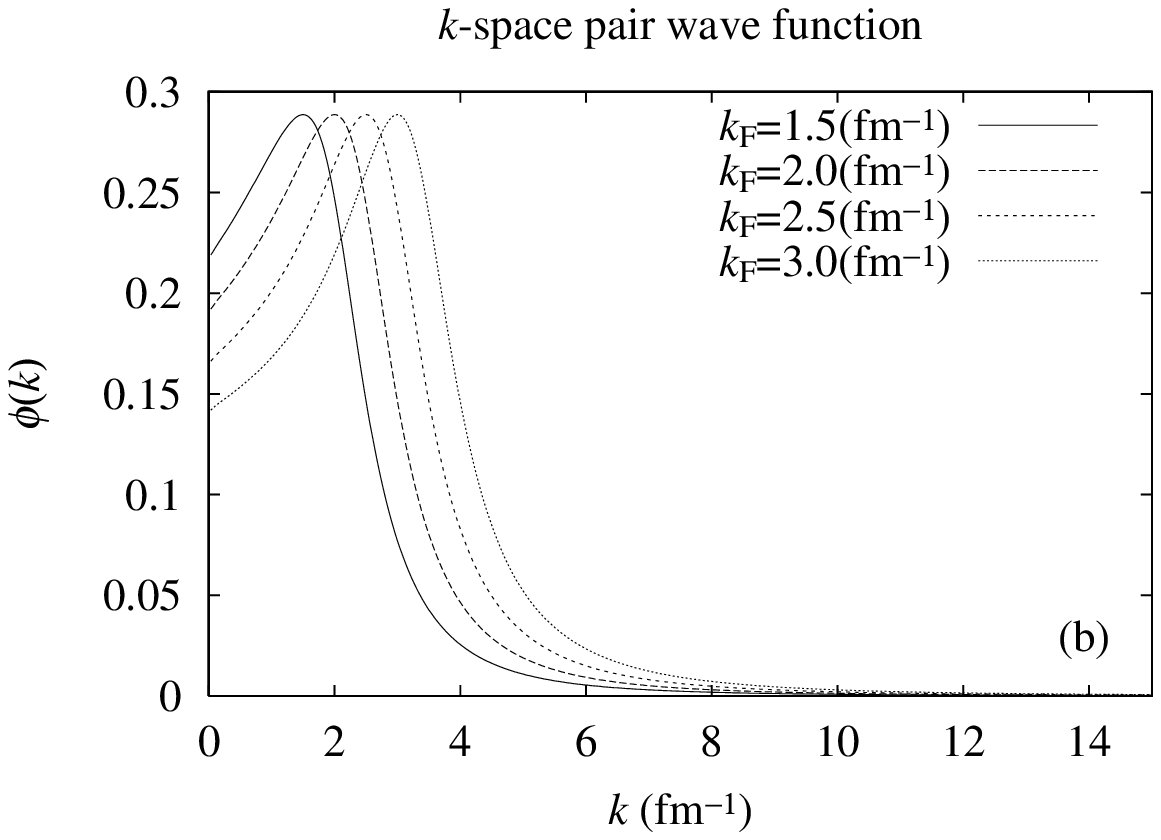,width=8cm}
\vskip 0.5cm
\epsfig{figure=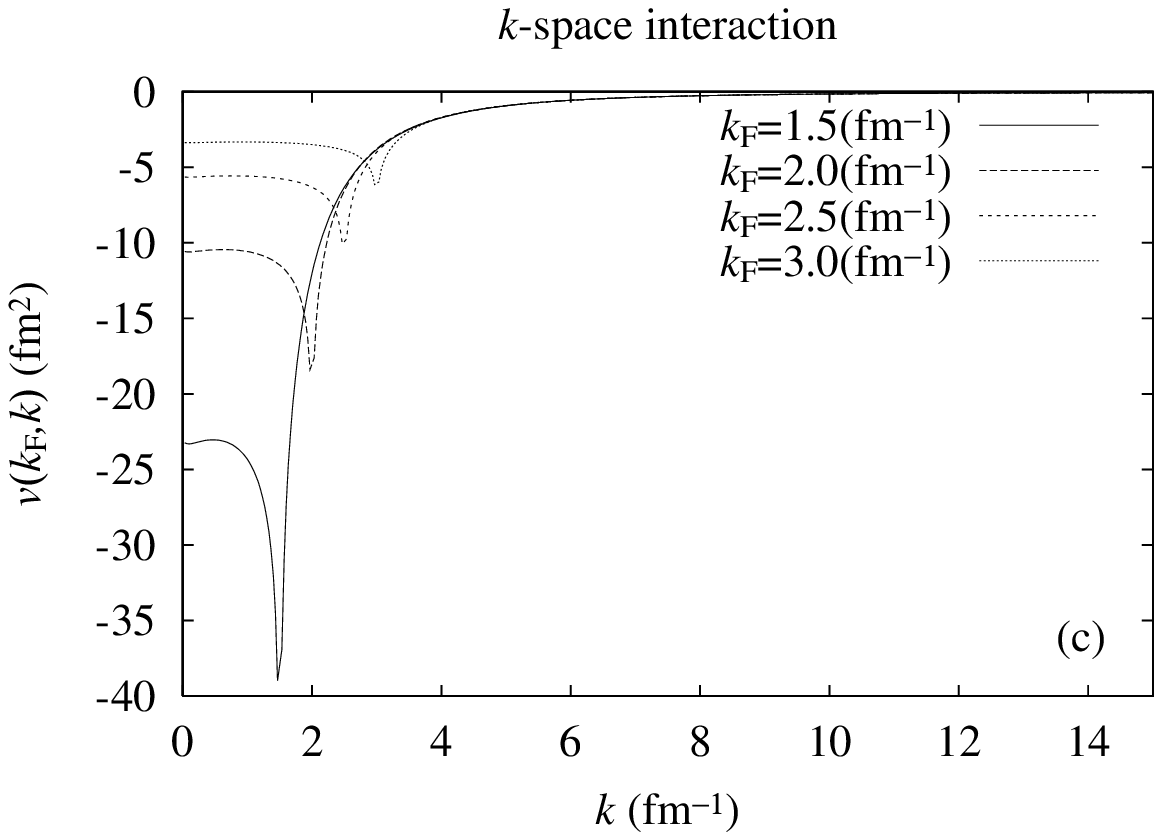,width=8cm}
\end{center}
\caption{
(a) Gap function, (b) pair wave function, 
and (c) matrix element $\bar v(k_{\rm F},k)$, as functions 
of the momentum $k$, calculated at $k_{\rm F}=$ 1.5 -- 3.0 
fm$^{-1}$. Note that, in (c), $k=k_{\rm F}$ is excluded 
since at this point the diverging magnetic interaction does not 
contribute to the gap equation [38].
}
\label{fig2}
\end{figure}

  In the following, we concentrate on the static case and present 
numerical results in three steps. 
As for the quark mass, $M_q=$ 10 MeV is adopted 
according to Ref.\cite{ii}. First we discuss the gap at the 
Fermi surface $\Delta(k_{\rm F})$ and the coherence length $\xi$ 
as functions of the Fermi momentum $k_{\rm F}$, which is related 
to the baryon density as 
$\rho=2k_{\rm F}^3/3\pi^2$ in the present 2 flavor case. 
The coherence length is defined as \cite{coh}, 
\begin{equation}
\xi=\left(\frac{\int_0^\infty\vert\frac{d\phi}{dk}\vert^2k^2dk}
               {\int_0^\infty\vert\phi\vert^2k^2dk}\right)
      ^{1/2}\, ,
\end{equation}
in the strong coupling case, in terms of the pair wave function, 
\begin{equation}
\phi(k)=\frac{1}{2}\frac{\Delta(k)}{E'(k)}
\, ,\label{eq7}
\end{equation}
which is identical to the pair condensate up to phase 
factors. In Fig.\ref{fig1}(a), $\Delta(k_{\rm F})$ is graphed 
for $k_{\rm F}=$ 1.5 -- 3.25 fm$^{-1}$, which corresponds 
to $\rho/\rho_0\simeq$ 1.5 -- 15 if the normal density of 
symmetric ($N=Z$) nuclear matter is defined as 
$\rho_0=2(k_{\rm F})_0^3/3\pi^2$, $(k_{\rm F})_0=$ 1.30 
fm$^{-1}$ \cite{sw2}. This shows very weak 
$k_{\rm F}$-dependence; the superconducting phase survives up to 
practically infinite density; for example, 
$\Delta(k_{\rm F}=60\,{\rm fm}^{-1})\simeq$ 50 MeV. 
We confirmed that the gap is 
predominantly brought about by the magnetic interaction; 
about 50\% of the total gap remains when the electric interaction 
is cut artificially while only 5 -- 10\% remains when the 
magnetic part is cut. This magnetic dominance can also be 
deduced from the strong $k_{\rm F}$-dependence of 
$\bar v(k_{\rm F},k)$ (Fig.\ref{fig2}(c) below) except at 
$k\simeq k_{\rm F}$ where the magnetic interaction gives 
very strong attraction irrespective of $k_{\rm F}$, in contrast 
to the weak $k_{\rm F}$-dependence of $\Delta(k_{\rm F})$. 
The coherence length is shown in Fig.\ref{fig1}(b). 
This can be compared with the Pippard length in the weak 
coupling theory, $\xi_0=k_{\rm F}/\pi\Delta(k_{\rm F})\mu
\simeq1/\pi\Delta(k_{\rm F})$ (see for example, 
Ref.\cite{degen}); the magnitudes of $\xi$ are approximated 
fairly well by $\xi_0$ whereas the $k_{\rm F}$-dependence 
is different. The average interparticle distance 
$d=\left({\frac{\pi^2}{2}}\right)^{1/3}/k_{\rm F}$, 
derived from $\rho_q=3\rho=1/d^3$, is also shown 
in the figure. The magnitudes of $\xi$ and $d$ are very 
similar to each other as in the nucleon-nucleon 
case \cite{tm}, whereas $\xi$ is 3 -- 4 orders of magnitude 
larger than $d$ in metals. This indicates strong coupling 
feature in the sense that quark Cooper pairs are compact 
and therefore bosonic. That is, fermion exchange is less when 
their mutual overlap is small. 
If $M_q$ is changed to 100 MeV and 300 MeV, 
for example, $\Delta(k_{\rm F})$ decreases to 90 -- 98 \% and 
54 -- 81 \% of the original values, respectively, for 
$k_{\rm F}=$ 1.5 -- 3.25 fm$^{-1}$, 
whereas the changes in $\xi$ are negligibly small. 

  Another quantity which can be compared with $\xi$ is the 
London penetration depth $\lambda_{\rm L}$, which is defined 
by $\sqrt{\mu/4\pi e_{\rm s}^2\rho_{\rm s}}$ in the 
relativistic case, where $e_{\rm s}$ and $\rho_{\rm s}$ are 
the electric charge of the Cooper pair and the density of the 
superconducting component, respectively. This expression 
can be derived from the relativistic Ginzburg-Landau theory 
developed in Ref.\cite{bl} (see also Refs.\cite{blas1,csc25}). 
In the present $ud$ 2 flavor case, 
$\lambda_{\rm L}\simeq$ 20 -- 8 fm for 
$k_{\rm F}=$ 1.5 -- 3.25 fm$^{-1}$ is obtained by adopting 
$e_{\rm s}=\frac{2}{3}e+(-\frac{1}{3})e$ and 
$\rho_{\rm s}=\frac{2}{3}\rho_q$. Although Bailin and Love 
concluded that quark matter is a superconductor of the 
first kind, 
$\xi>\sqrt{2}\lambda_{\rm L}$, based on their estimate, 
$\Delta(k_{\rm F})\sim$ 1 MeV \cite{bl}, the present 
quantitative study shows that quark matter is a 
superconductor of the second kind at these densities as 
pointed out in Ref.\cite{blas1}.
At asymptotically high densities, however, it changes to 
a superconductor of the first kind since the penetration depth 
decreases rapidly as density increases.
One should note that these discussion applies only to a part 
of magnetic field since the original photon and a gluon combine 
to a massless `rotated' photon although both the electromagnetic 
U(1) and the color SU(3) break in color superconductors as recently 
discussed in Ref.\cite{rotphot}.

  Next, we turn to $k$-dependence at each $k_{\rm F}$. 
The compactness of the Cooper pairs mentioned above indicates spreading 
of $\Delta(k)$ (Fig.\ref{fig2}(a)) and $\phi(k)$ (Fig.\ref{fig2}(b)) in 
$k$-space. Figure \ref{fig2}(a) shows that $\Delta(k)$ 
spreads to larger $k$ at higher densities as expected, while 
the width of $\phi(k)$ in Fig.\ref{fig2}(b), which corresponds to 
$1/\xi$, is almost $k_{\rm F}$-independent as mentioned above 
since the quasiparticle energy $E'(k)$ in the denominator in 
Eq.(\ref{eq7}) grows as $k$ goes away from $k_{\rm F}$. 
The asymmetric shapes of $\phi(k)$ reflect the smallness of $\xi$. 
Its width is similar to that of $\bar v(k_{\rm F},k)$ in 
Fig.\ref{fig2}(c).

  Finally, we Fourier-transform $\phi(k)$ to
\begin{equation}
\phi(r)=\frac{1}{2\pi^2}\int_0^\infty\phi(k)j_0(kr)k^2dk\, ,
\end{equation}
where $j_0(kr)$ is a spherical Bessel function, in order to 
look into the spatial structure of quark Cooper pairs more 
closely. This quantity has already been discussed for nucleon 
Cooper pairs in non-relativistic \cite{baldo,nrel4} and 
relativistic \cite{tm} studies. Although the coherence length 
$\xi$, corresponding to the root mean square distance with 
respect to this wave function, is almost $k_{\rm F}$-independent, 
the shape of $\phi(r)$ is strongly $k_{\rm F}$-dependent; 
large-$k$ components in high-density cases bring nodes in 
$r$-space. Conversely, $\phi(r)$ is diffuse at low densities. 
In other words, quark Cooper pairs become more bosonic at 
higher densities.

  To summarize, we have studied numerically the spatial structure 
of quark Cooper pairs in the color ${\bf 3}^{\ast}$, 
isosinglet $^1S_0$ channel by solving the gap equation in full 
momentum range. Although the long-range magnetic interaction 
is predominantly responsible for the quark-quark pairing, the gap 
function spreads also in momentum space. The resulting coherence 
length is almost density independent as well as the gap at the 
Fermi surface and is of magnitudes similar to the average 
interquark distance. The dependence of the pair wave function 
(condensate) on the relative momentum and on the 
distance between two quarks that form a Cooper pair has been presented. 
This indicates that quark Cooper pairs become more bosonic 
at higher densities although the coherence length is almost 
density independent. 

  The author thanks T. Hatsuda for informing him of R. Horie's 
master thesis in Ref.\cite{hh} and for the communication of 
Ref.\cite{hat}.

\newpage
\newpage
\newpage
\begin{figure}[t]
\begin{center}
\epsfig{figure=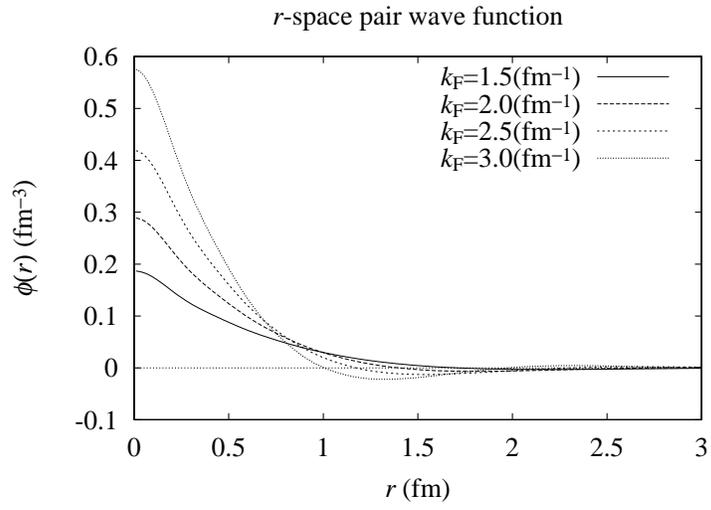,width=10cm}
\end{center}
\caption{
Pair wave function as a function of the distance $r$, 
calculated at $k_{\rm F}=$ 1.5 -- 3.0 fm$^{-1}$.
}
\label{fig3}
\end{figure}

\end{document}